\documentclass[10pt]{iopart}
\usepackage{iopams}
\usepackage{epsfig}

\newcommand{\sgn}{\mbox{sgn}}
\newcommand{\erf}{\mbox{erf}}

\newcommand{\bw}{\ensuremath{\mathbf{w}}}
\newcommand{\bx}{\ensuremath{\mathbf{x}}}
\newcommand{\by}{\ensuremath{\mathbf{y}}}

\newcommand{\RR}{\ensuremath{\mathbb{R}}}

\newcommand{\II}{\ensuremath{\mathbb{I}}}
\newcommand{\PP}{\ensuremath{\mathbb{P}}}

\newcommand{\hc}{\ensuremath{\widehat{c}}}

\newcommand{\hr}{\ensuremath{\widehat{r}}}

\newcommand{\hw}{\ensuremath{\widehat{w}}}
\newcommand{\hx}{\ensuremath{\widehat{x}}}
\newcommand{\hy}{\ensuremath{\widehat{y}}}

\newcommand{\hC}{\ensuremath{\widehat{C}}}
\newcommand{\hG}{\ensuremath{\widehat{G}}}

\newcommand{\hK}{\ensuremath{\widehat{K}}}
\newcommand{\hR}{\ensuremath{\widehat{R}}}
\newcommand{\hW}{\ensuremath{\widehat{W}}}
\newcommand{\hsigma}{\widehat{\sigma}}

\newcommand{\hbw}{\ensuremath{\widehat{\mathbf{w}}}}
\newcommand{\hbx}{\ensuremath{\widehat{\mathbf{x}}}}
\newcommand{\hby}{\ensuremath{\widehat{\mathbf{y}}}}
\newcommand{\hbsigma}{\widehat{\bsigma}}

\begin{document}

\title[Supervised Learning with Restricted Training Sets]{Supervised
Learning with Restricted Training Sets: a Generating
Functional Analysis}
\author{J A F Heimel and A C C Coolen}
\address{Department of Mathematics, King's College London, The Strand,
London WC2R~2LS, UK}
\begin{abstract}
We study the dynamics of supervised on-line learning of realizable tasks
in feed-forward neural networks. We focus on the regime where the
number of examples used for training is proportional to
the number of input channels $N$. Using  generating function
techniques from spin glass theory, we are able to average over the
composition of the training set and transform the problem for
$N\rightarrow\infty$ to an effective single pattern system, described
completely by the student autocovariance, the student-teacher overlap
and the student response function, with exact closed 
equations. Our method applies to arbitrary learning rules, i.e. not
necessarily of a gradient-descent type.
The resulting exact macroscopic 
dynamical equations can be integrated without
finite-size effects up to any degree of accuracy, but their main value
is in providing an exact and simple starting point for analytical
approximation schemes. Finally,  we show how, in the region of absent
anomalous response and using the hypothesis that (as in detailed
balance systems) the short-time part of the various operators can be
transformed away, one can describe the stationary state of the
network succesfully by a set of coupled equations involving 
only four scalar order parameters.
\end{abstract}



\section{Introduction}
It is now a little more than ten years since studies of the dynamics
of supervised learning in artificial neural networks started appearing
in the statistical physics literature.  Early theoretical studies
focussed on on-line learning using complete training sets where the
probability of the same example appearing twice during training was
zero, e.g. \cite{KinzRuja90,BiehSchw92,BiehRieg94}. This work enabled
the evaluation of properties like convergence speed, generalization
ability and optimal learning rates.  However, such studies were still
significantly removed from real-world scenarios. The most serious
restriction was that one had to assume the availability of an infinite
amount of training date, homogeneously distributed over the input
space.  In a recent article \cite{RaeHeimCool00} it was shown that
even for very simple inhomogenuity the generalization error is no
longer self-averaging and deterministic. The issue of repeating
examples during training is technically a much harder problem and has
received much attention recently. Most of the work has focussed on
simple or linear learning rules
\cite{HertKrogThor89,RaeSollCool99,SollBarb98} or different kinds of
approximations, such as Fokker-Planck approaches \cite{HeskKapp91,
HansPathSala93,Rado93,Hesk94} and Gaussian local field distributions
\cite{BarbSoll98}. Exact work on non-linear learning rules has drawn
heavily on techniques from the spin glass and disordered systems
community (for an early overview of these techniques see
e.g. \cite{MezaPariVira87}).  The generating functional technique was
used to study the dynamics of Gibbs learning in a perceptron with
binary weights in \cite{Horn92a,Horn92b}. A dynamical version of the
cavity method was employed in \cite{LiWong00,WongLiTong00,WongLiLuo00}
to study gradient descent batch learning and the methods of dynamical
replica theory were applied to the problem of on-line learning
in \cite{CoolSaad98,CoolSaad00,CoolSaadXion00,MaceCool00}.  The
on-line learning scenario in this last sequence of papers is the one
that we study here, but in the present paper we adapt the generating
functional method \`a la De Dominicis to deal with on-line learning.
This paper might be the first to present exact macroscopic equations
for on-line learning of restricted training sets for non-linear
learning rules which are not of a gradient-descent type.

Precise definitions will be given in section
\ref{sec:definitions}, but the general setup is the following. The
examples presented to the student perceptron are $N$ dimensional
vectors chosen with equal probability from a fixed training set
$\Omega$. The number of examples in $\Omega$ is $p=\alpha N$. At each
presentation the student is given the teacher's classification of the
pattern. The student can then decide to change its `program',
represented by the $N$ dimensional vector $\bsigma \in \RR^N$, in order to
 resemble more the teacher's program $\btau\in\RR^N$. The random choice of
a pattern from the training set makes the evolution of the student
weight vector $\bsigma$ a stochastic process. In section
\ref{sec:genfun} we write down a generating function for all the
possible paths of $\bsigma$. This function can be averaged over all
possible realizations of the training set $\Omega$ (a quenched
disorder average). At that point we will take the limit $N$ to
infinity, to find saddle-point equations for a set of five order
parameters and their conjugates. The reader who is mainly interested
in results can skip section \ref{sec:genfun} and go directly to
 section
\ref{sec:eff}, where the equations are reduced to a single exact set
of three equations involving the student autocorrelation
$C(t,t')=\bsigma(t)\cdot\bsigma(t')/N$, the student-teacher overlap
$R(t)=\bsigma(t)\cdot\btau/N$ and the student response function
$G(t,t')$.  This set gives a surprisingly simple and intuitive picture
of the evolution of the order parameters and the distribution of the
local fields. From that point it is easy to establish links with
earlier work on infinite training sets, batch learning and linear
learning rules. Numerical evidence is presented, showing that the
present theory is in very good agreement with the simulations. 

In section \ref{sec:statstate}, the stationary state of a student with
constant weight decay is studied. For the stationary state one can
split all relevant order parameters into persistent and non-persistent
parts. If we keep only the persistent parts and the single-time
non-persistent parts, we find a closed set of equations containing
just four scalar order parameters. The procedure is inspired by a
similar method applied to the solution of detailed balance 
spin glass dynamics, where
it can be shown to be exact. Although the numerical evidence certainly
seems to suggest that the procedure yields the correct results, we can
not proof this fact rigorously here. At the moment, it remains an
interesting open question.


\section{Definitions}\label{sec:definitions}

We study on-line learning in a student perceptron characterized by a
vector $\bsigma\in\RR^N$. The student classifies
patterns $\bxi \in \Omega \subset \{-1,+1\}^N$ according to
$S(\bxi)=\sgn(\bsigma\cdot \bxi)$. The student tries to learn the task
set by the teacher $T(\bxi)=\sgn(\btau\cdot\bxi)$ with
$\btau\in\RR^N$,
 i.e. we only
consider linear separable classifications. The components of
the weight vectors of teacher and student are assumed not to scale with
$N$. The set $\Omega$ contains only
$p=\alpha N$ examples, independently chosen with equal probability from
$\{-1,+1\}^N$. Patterns will be labeled by the Greek index $\mu$.
At each iteration each pattern is equally likely to be
chosen for presentation to the student,
 independently of previous rounds. If at
step $m$, pattern $\mu(m)$ is presented to the learning student, the
student's  weight vector is slightly adjusted to converge to the desired
classification according to a recipe of the general form:
\begin{equation}
   \bsigma(m+1)=\bsigma(m)+\frac{\eta}{\sqrt{N}}\bxi^{\mu(m)}
    F\left(
      \frac{\bsigma(m)\cdot\bxi^{\mu(m)}}{\sqrt{N}},
      \frac{\btau\cdot\bxi^{\mu(m)}}{\sqrt{N}}
    \right).
\end{equation}
The speed of the evolution is set by the learning rate $\eta$. 
The function $F(x,y)$ is the
learning rule. Popular learning rules are e.g.
\begin{equation}
  F(x,y)=\left\{\begin{array}{ll}
    y-x,& \mbox{Linear} \\ 
    \sgn(y), & \mbox{Hebb} \\
    \sgn(y)-x, & \mbox{Adaline} \\
    \sgn(y)\theta(-xy), &  \mbox{Perceptron} \\
    |x|\sgn(y)\theta(-xy), & \mbox{Adatron} 
  \end{array}\right.
\end{equation}
where $\theta$ is the stepfunction, $\theta(x)=1$ for $x\geq
0$ and $\theta(x)=0$ for $x<0$. The first  three learning
rules are all linear in $x$, while the last two only  alter the
student's weights
when student and teacher disagree.

A theoretical study of perceptrons can be useful for predicting learning
times, for evaluating different learning rules or for finding optimal learning
rates. For this purpose one is not so much interested in predicting the
specific microscopic realizations of $\bsigma$ over time, but rather
in the number of errors the perceptron makes in the classification of
the training set (training error, $E_t$) and the number of errors 
in the classification 
of the complete set of
examples $\{-1,+1\}^N$ (generalization error, $E_g$):
\begin{eqnarray}
  E_t(\bsigma) & 
  \equiv & 
  \langle  \theta( - (\bsigma\cdot\bxi)(\btau\cdot\bxi)) \rangle_\Omega
  =
  \frac{1}{p}\sum_{\bxi\in\Omega} 
    \theta( - (\bsigma\cdot\bxi)(\btau\cdot\bxi)),
  \\
  E_g(\bsigma) &
  \equiv & 
  \langle   \theta( - (\bsigma\cdot\bxi)(\btau\cdot\bxi)) \rangle
  =
  \frac{1}{2^N}\sum_{\bxi\in\{-1,+1\}^N}
    \theta( - (\bsigma\cdot\bxi)(\btau\cdot\bxi)).
\end{eqnarray}
Given $\bsigma$, the generalization error is independent
of the training set.
It is in fact a standard result in perceptron theory that this error
is only dependent on the angle between student and teacher vector, i.e. 
the norm of $\bsigma$ and its overlap with $\btau$.
\begin{equation}
 E_g(\bsigma)=\frac{1}{\pi}\arccos\left(
  \frac{R(\bsigma)}{\sqrt{C(\bsigma,\bsigma)}}
 \right).
\end{equation}

\section{The generating functional}\label{sec:genfun}

The random choice of a pattern $\mu(m)$ makes it more convenient to go
to a description of an ensemble of students with a
distribution of weight vectors, $P_m(\bsigma)$, than to study the
stochastic evolution of $\bsigma$ directly. 
In this setting we can study the (moment) generating function $Z_M$
for iteration times up to $M$:
\begin{equation} 
  Z_M[\bpsi]
 =
 \int \!D\bsigma\, P(\bsigma(0),\bsigma(1),\ldots,\bsigma(M))
   \e^{i \sum_{m=0}^M \bpsi(m)\cdot\bsigma(m)},
\end{equation}
where $\int D_{m=0}^M \bsigma=\int \prod_m d\bsigma(m)$ is an integral over all
possible paths the students could take. 
Derivation of $Z_M$ with respect to $\bpsi$ generates all moments of
the distribution $P$.
The microscopic dynamics of weight vectors at time $m$ can be
written in the general form
$P_{m+1}(\bsigma)=\int d\bsigma' W(\bsigma | \bsigma')P_m(\bsigma')$,
with the transition probabilities 
\begin{equation}
  W(\bsigma|\bsigma')
  =
  \frac{1}{p}\sum_{\mu=1}^p
    \delta\left(
      \bsigma-\bsigma'-\frac{\eta}{\sqrt{N}}\bxi^\mu
      F\left(
        \frac{\bsigma'\cdot\bxi^\mu}{\sqrt{N}},
        \frac{\btau\cdot\bxi^\mu}{\sqrt{N}}
      \right)
   \right)
\end{equation}
To disentangle the double $\bxi$ dependence of the transition rates,
we employ the integral representation
 of the Dirac delta-function and introduce
\begin{eqnarray}
  W(\bsigma|\bsigma',\bx',\by)
  &=&
  \frac{1}{p}\sum_\mu
    \delta\left(
     \bsigma-\bsigma'-\frac{\eta}{\sqrt{N}}\bxi^\mu F(\bx'^\mu,y^\mu)
   \right)
  \\
  &=&
  \int \frac{d\widehat\bsigma}{(2\pi)^N}
  \exp \left[i\widehat{\bsigma}\cdot(\bsigma-\bsigma')\right]
  \widehat{W}(\hbsigma | \bx',\by,\bw ),
\end{eqnarray}
where we introduced three shorthands, called local fields (in
analogy with spin systems)
\begin{equation}
x^\mu=\frac{1}{\sqrt{N}}\bsigma\cdot\bxi^\mu,\quad
y^\mu=\frac{1}{\sqrt{N}}\btau\cdot\bxi^\mu, \quad
w^\mu=\frac{1}{\sqrt{N}}\hbsigma\cdot\bxi^\mu,
\end{equation}
and the Fourier transform of the transition rate $\hW$. For large $N$,
$\hW$ will be of order
$1+\mathcal{O}(N^{-1/2})$ and will therefore factorize over the patterns
\begin{eqnarray}
  \widehat{W}(\hbsigma | \bx,\by,\bw )
  &=&
  \frac{1}{p}\sum_\mu \exp\left(
     -i \eta w^\mu  F(x^\mu,y^\mu)
   \right)
  \\
  &=&
    \exp\left[\frac{1}{p}\sum_\mu
    \e^{-i\eta w^\mu F(x^\mu,y^\mu)} -1\right],
   \quad (N\rightarrow\infty) \nonumber
\end{eqnarray}
We can now rewrite the generating function:
\begin{eqnarray*}
\fl
  Z_M[\bpsi]
  =
  \int  D\bsigma P_0(\bsigma(0)) \prod_{m=0}^{M-1}W(\bsigma(m+1)|\bsigma(m))
  \\
  \lo =\!
  \int\!\!\! \frac{D\bsigma D\hbsigma}{(2\pi)^N} D\bx d\by D\bw 
   P_0(\bsigma(0))
   \Gamma[\by,\bx,\bw,\bsigma] \prod_m \hW(\hbsigma(m)|\bx(m),\by,\bw(m))  \\
  \times
  \prod_m
  \exp\left[i
   \hbsigma(m)\cdot(\bsigma(m+1)-\bsigma(m)) +i\bpsi(m)\cdot\bsigma(m) \right]
\end{eqnarray*}
where the appearance of the training examples is restricted to the
function $\Gamma$, given by:
\begin{eqnarray*}
\fl  \Gamma[\by,\bx,\bw,\bsigma]
  &=&
  \prod_\mu \delta\left[y^\mu-\frac{\btau\cdot\bxi^\mu}{\sqrt{N}}\right] 
  \prod_m \delta\left[x^\mu(m)-\frac{\bsigma(m)\cdot\bxi^\mu}{\sqrt{N}}\right] 
  \delta\left[w^\mu(m)-\frac{\hbsigma(m)\cdot\bxi^\mu}{\sqrt{N}}\right] 
\end{eqnarray*}
In the thermodynamic limit $(N\rightarrow\infty$), all the macroscopic observables
in this model are self-averaging with respect to the realization of
the training set. 
To avoid the difficulty of choosing a typical training set, we can
thus safely consider the disorder averaged generating function
$[ Z ]_{dis}$. The only term involving the actual patterns is
$\Gamma$. The quenched disorder average of $\Gamma$ is
\begin{eqnarray*}
\fl \left[\Gamma\right]_{dis}
  =  
  \int d\hby D\hbx D\hbw \prod_\mu
  \exp\left[
    i \hy^\mu y^\mu +\sum_m i \hx^\mu(m)x^\mu(m)+ \sum_m i \hw^\mu(m)w^\mu(m)
  \right]
  \\ 
 \times 2^{-N}\!\!\!\!\!\!\!\sum_{\bxi^\mu \in \{\pm1\}^N}\!\!\!\!\!\!\!
  \exp\frac{-\rmi}{\sqrt{N}}\bxi^\mu \!\cdot \!\left[
     \hy^\mu \btau
    +\sum_m \hx^\mu(m)\bsigma(m)
    +\sum_m \!\hw^\mu(m)\hbsigma(m)
  \right]
\end{eqnarray*}
Of the term on the second line, only the quadratic terms in
 $\btau$, $\bsigma$ and $\hbsigma$ survive in the thermodynamic
 limit. Near this limit we find that this term containing the training
 patterns becomes
\begin{eqnarray*} 
\prod_{\mu,i} 
  \exp\left[-\frac{1}{2N}
    \left( 
       \hy^\mu \tau_i
       +\sum_m \hx^\mu(m)\sigma_{i}(m)
       +\sum_m \hw^\mu(m) \hsigma_{i}(m)
    \right)^2
  \right],
\end{eqnarray*}
in the thermodynamic limit.
We assume that the initial probability distribution $P_0(\bsigma)$
factorizes over sites.
Full factorization of the generating function over patterns and input
channels can then be achieved if we introduce the following order
parameters and their conjugates via delta-functions:
\begin{eqnarray*}
  R_m   =   \frac{1}{N} \sum_i \sigma_{i}(m) \tau_i, 
  & r_m  =  \frac{1}{N} \sum_i\hsigma_{i}(m) \tau_i, \\
  C_{mn}  =  \frac{1}{N} \sum_i \sigma_{i}(m) \sigma_{i}(n),\qquad &
  c_{mn}  =  \frac{1}{N} \sum_i\hsigma_{i}(m)\hsigma_{i}(n), \\
  K_{mn}  =  \frac{1}{N} \sum_i\sigma_{i}(m) \hsigma_{i}(n). 
\end{eqnarray*}
When changing $m$ to $m+1$, the expectation of these order parameters can
only can by a value of order $N^{-1}$. 
We thus rescale the time as 
$t= m/\Delta N$. From here, one could go to a continuous time
description by introducing $\tau=\Delta t$ and taking the limit 
$\Delta$ to zero, but we delay this step in order to avoid technical
difficulties in evaluating the path integrals.
The generating function attains a form suitable for saddle-point
integration:
\begin{equation}
  \left[Z[\bpsi]\right]_{dis}
  \propto
  \int \ldots
  \exp \left[N (\Psi +\Phi+\Omega) \right]
\end{equation}
There are three distinct leading order contributions to the
exponent. The first is a `bookkeeping' term, linking the order
parameters to their conjugates:
\begin{equation}
  \Psi=i\hR\cdot R + i\hr\cdot r  + i\Tr\left[
    \hC^T C + \hK^T K + \hc^T c
  \right]
\end{equation}
The second term reflects the coupled dynamics of the local
fields:
\begin{eqnarray}
\fl  \label{eq:Phi}
  \Phi=\frac{1}{N}\sum_\mu \log \int
  \frac{dyd\hy}{2\pi}\frac{DxD\hx}{(2\pi)^T}\frac{D\hw Dw}{(2\pi)^T}\exp\left[
    i\hw\cdot w 
    + \frac{\Delta}{\alpha}\sum_t \left(\e^{-i\eta w_t F(x_t,y)}-1 \right)
  \right. \nonumber
\\ \left.
  +  i\hy(y-\theta^\mu_y) + i\hx\cdot (x-\theta^\mu_x) 
  -\frac{1}{2}\hx C\hx
  -\frac{1}{2}\hy^2
  \right.
\\\left.
   - \hx K \hw  - \hy R\cdot\hx
    -\frac{1}{2}\hw c \hw - \hy r\cdot\hw 
  \right],  \nonumber
\end{eqnarray}
where we have added additional sources $\btheta_x$ and $\btheta_y$ to
couple to $\hbx$ and $\hby$. These sources act as biases of teacher
and student.
The third term describes the evolution of the now decoupled weight
components:
\begin{eqnarray} \fl
  \Omega 
  =
  \frac{1}{N}\sum_i \log\int \frac{D\sigma D\hsigma}{(2\pi)^T}
  P_{0i}(\sigma_0)   \label{eq:Omega}
  \exp 
  \left[
  -i\tau_i \hR\cdot \sigma
  -i\tau_i \hr \cdot \hsigma
\right.
\\
\left.
  -i\sigma \hC \sigma
  -i\hsigma \hc \hsigma
  -i\sigma \hK \hsigma
  +i\hsigma G_0^{-1} \sigma -i\hsigma \cdot \theta_i + i\sigma \cdot \psi_i
  \right] \nonumber
\end{eqnarray}
where $[G_0^{-1}]_{tt'}=\delta_{t+1,t'}-\delta_{tt'}$ and
 where we have included an external driving force $\theta_{i}(t)$ in the
system. With a modest amount of foresight we write
$G_{tt'}=-iK_{tt'}$. Upon taking derivatives with respect to the
generating fields $\{\psi_{i}(t),\theta_{i}(t)\}$, we find \emph{at} the relevant
saddle-point:
\begin{eqnarray*}   
  R_t=\lim_{N\rightarrow\infty}
        \frac{1}{N}\sum_i 
          \left[ \langle \sigma_{i}(t)\tau_i\rangle
       \right]_{dis},
\\
  C_{tt'}=\lim_{N\rightarrow\infty}\frac{1}{N}\sum_i 
    \left[\langle \sigma_{i}(t)\sigma_{i}(t')\rangle \right]_{dis},
\\
  G_{tt'}=\lim_{N\rightarrow\infty}\frac{1}{N}\sum_i 
     \frac{\partial}{\partial \theta_i(t') }
     \left[\langle \sigma_{i}(t) \rangle\right]_{dis}
\end{eqnarray*}
Using the built-in
normalisation $\left[Z(0)\right]_{dis}$, we also find
\begin{eqnarray*}
  r_{t}=\lim_{N\rightarrow\infty}\frac{1}{N}\sum_i 
        \frac{\partial}{\partial \theta_{i}(t) }
     \left[\langle \tau_i \rangle\right]_{dis}=0,
\\
  c_{tt'}=\lim_{N\rightarrow\infty}\frac{1}{N}\sum_i
     \frac{\partial^2}{\partial
     \theta_{i}(t)  \partial \theta_{i}(t')}
     \left[Z(0)\right]_{dis}=0
\end{eqnarray*}
If we perform the saddle-point integration, we find in addition that
\begin{eqnarray*}
  i\hR_t=-\lim_{N\rightarrow\infty}\frac{1}{N}\sum_\mu
    \frac{\partial^2}{\partial \theta^\mu_y \partial \theta^\mu_{x}(t)}
    \left[Z(0)\right]_{dis}=0,
\\
  i\hC_{tt'}=- \lim_{N\rightarrow\infty} \frac{1}{N}\sum_\mu
    \frac{\partial^2}{\partial \theta_{x}(t) \partial \theta_{x}(t')}
    \left[Z(0)\right]_{dis}=0.
\end{eqnarray*}
At this point we can already simplify (or remove altogether)
 the generating fields $\theta_i(t)=\theta_t, 
\theta_x^\mu(t)=\theta_{xt},
\theta_y^\mu(t)=\theta_{yt}$ and $\psi_i(t)=0$. 
The external fields $\theta_x$ and $\theta_y$ can be interpreted as
biases or thresholds of the student and teacher, respectively.
Without
loss of generality we may
set $\tau_i=1$.
The evolution of the local fields and the weight vector
are now linked  only via the remaining non-zero order
parameters. We proceed to evaluate the two separate processes at the saddle-point.

\subsection{Pattern average $\Phi$}
Focussing on the evaluation of the pattern average $\Phi$ we find that
the terms involving $w$ can be interpreted as averages over a
Poisson-distribution:
\begin{eqnarray*} \fl
  \int \frac{dw_t}{2\pi} \exp\left[
    i\hw_t w_t + \frac{\Delta}{\alpha}\left(\e^{-i\eta w_t F(x_t,y)}-1\right)
  \right]
  \\
  =
  \sum_{k_t=0}^\infty \int \frac{dw_t}{2\pi} \exp\left[
    i\hw_t w_t -i\eta k_t w_t F(x_t,y) - \frac{\Delta}{\alpha}
  \right]
  \frac{1}{k_t!}\left(
    \frac{\Delta}{\alpha}
  \right)^{k_t}
  \\
  =\sum_{k_t=0}^\infty \delta(\hw_t-\eta k_t F(x_t,y)) \PP(k_t),
\end{eqnarray*}
where $\PP(k)$ is a Poisson distribution with average
$\Delta/\alpha$. For $\Delta N\gg 1$, $\PP(k)$ gives the probability
that a specific pattern is presented $k$ times to the student in time
interval $\Delta$. The saddle-point equations of the remaining
non-zero order parameters are found to be:
\begin{equation}\label{eq:hrhchG}
  \hr_t=\alpha\frac{\partial}{\partial \theta_y}
      \langle f_t \rangle_\Phi,\quad
  2i\hc_{tt'}=\alpha\langle f_t f_{t'} \rangle_\Phi,\quad
  i\hG_{tt'}=-\alpha \frac{\partial}{\partial \theta_{xt}}
    \langle f_{t'} \rangle_\Phi,
\end{equation}
with the shorthand $f_t=\eta k_t F(x_t,y)$. The average
$\langle\cdot\rangle_\Phi$ is using the measure implied by equation
(\ref{eq:Phi}). Performing the disorder average has turned the $\hy$ integral
into a Gaussian one. Evaluating this integration yields:
\begin{eqnarray}
  \Phi&=&\alpha \log \int \frac{dy}{\sqrt{2\pi}}\frac{DxD\hx}{(2\pi)^T} 
  \prod_t \left[\sum_{k_t}\PP(k_t) \right]
  \label{eq:Phi2}\\ 
&&  \exp\left[
    -\frac{1}{2}(y-\theta_y)^2 - \frac{1}{2}\hx D \hx 
    +i\hx\cdot(x-\theta_x- Gf - R(y-\theta_y))
  \right],\nonumber
\end{eqnarray}
where we have introduced the student autocovariance 
$D_{tt'}=C_{tt'}-R_tR_{t'}$. We note the 
operator identity $\partial/\partial \theta_y = y -
R\cdot\partial/\partial \theta_x$, which in turn implies using
(\ref{eq:hrhchG}) that
\begin{equation} \label{eq:hr}
  \hr_t= \alpha \langle yf_t\rangle_\Phi
     +\sum_{t'} i\hG^T_{tt'}R_{t'}
\end{equation}

\subsection{Weight component average $\Omega$}
The saddle-point equations involving the weight vectors are:
\begin{equation}
  R_t=\langle \sigma_t \rangle_\Omega,\qquad
  C_{tt'}=\langle \sigma_t \sigma_{t'}\rangle_\Omega,\qquad
  G_{tt'}=\frac{\partial}{\partial \theta_{t'}}\langle \sigma_t \rangle_\Omega,
\end{equation}
where $\langle\cdot\rangle_\Omega$ is an average with the measure
induced by (\ref{eq:Omega}).
This measure can be generated by the stochastic process:
$
   -\hr+i\hG^T\sigma+G^{-1}_0\sigma-\theta-\rho=0,
$
where $\rho_t$ is a Gaussian noise with zero mean and covariance
$\langle \rho_t \rho_{t'}\rangle=\Lambda_{tt'}\equiv 2i\hc_{tt'}$. 
From this process, we find a simple expression
for $\sigma$ (upon setting $\theta=0$):
\begin{equation}
  \sigma=G(\hr+\rho),
\end{equation}
with the response, student-teacher overlap and student autocovariance
given by
\begin{equation} \label{eq:GRD1}
  G=\left[G_0^{-1}+i\hG^T\right]^{-1}\!\!\!, \qquad
  R=G\hr,\qquad
  D=G\Lambda G^T
\end{equation}

\section{Effective single pattern process}\label{sec:eff}

\begin{figure}
  \begin{tabular}{cc}
     \epsfig{file=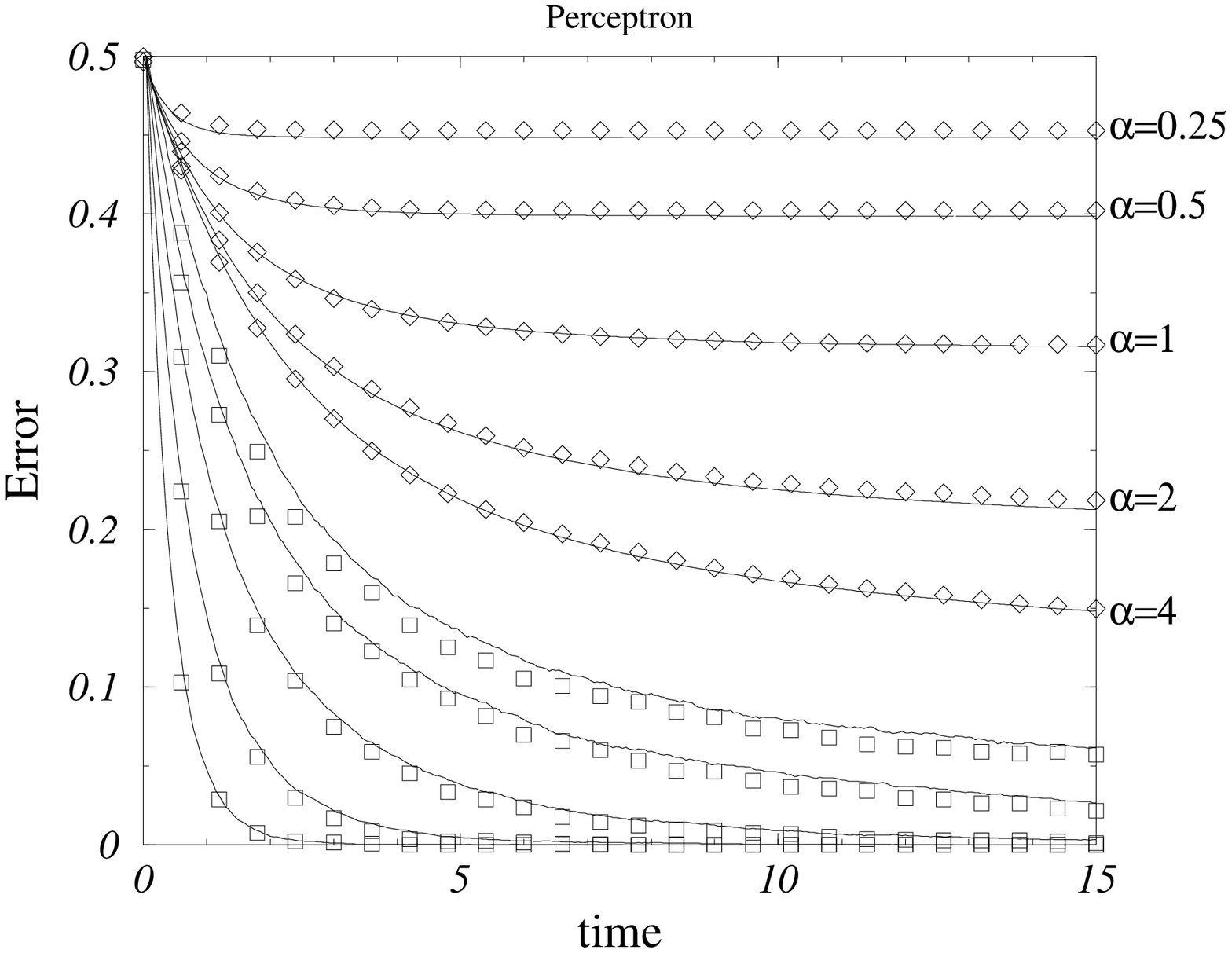,width=0.45\textwidth}
  &
     \epsfig{file=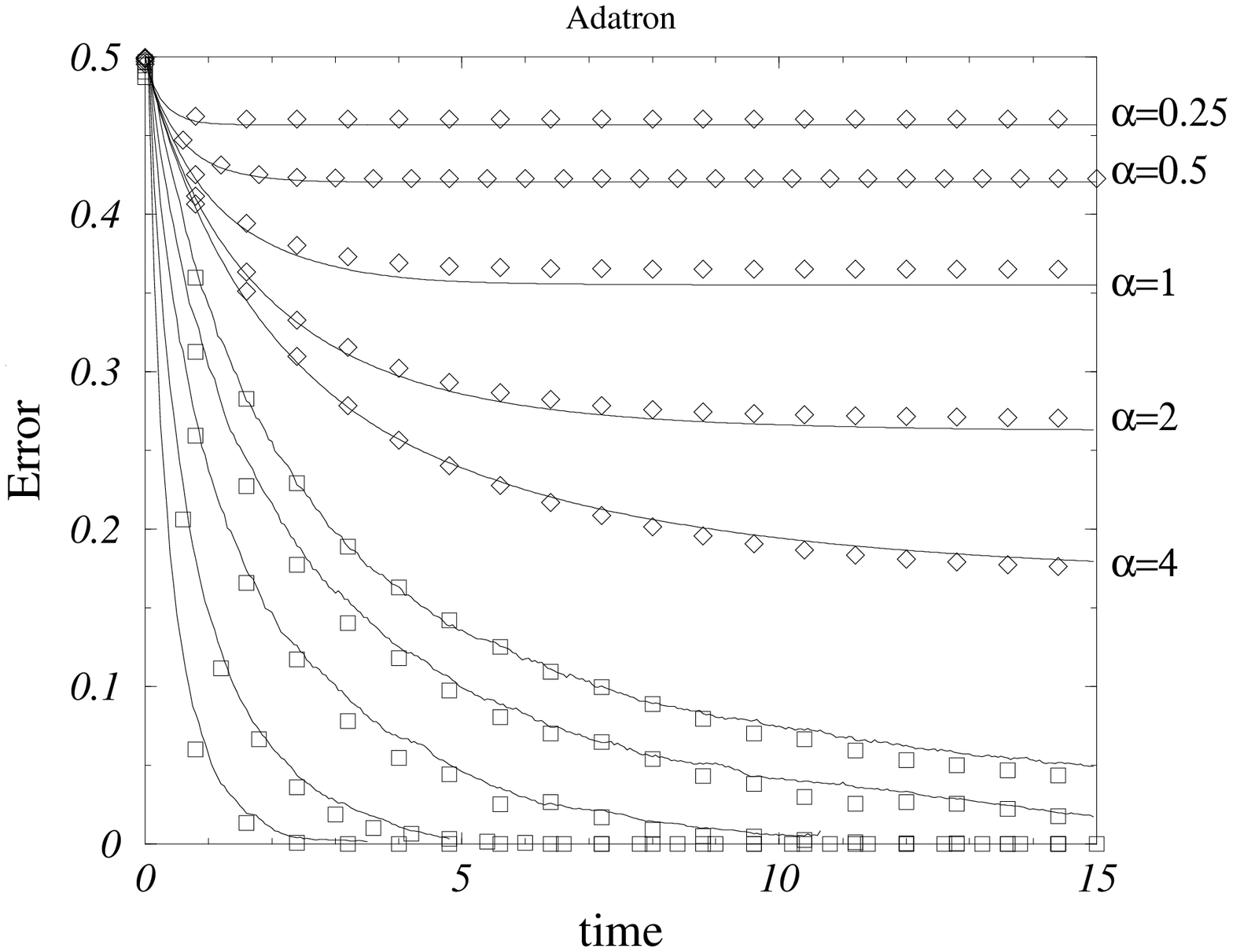,width=0.45\textwidth}
  \end{tabular}
\caption{\label{fig:evolution}The evolution of the generalization
(upper lines with diamonds) and training error (lower lines with
squares) for the perceptron (left) and
adatron (right) learning rules for various training set sizes. The
lines correspond to (generalization error: top to bottom, training
error: bottom to top) $\alpha=0.25,0.5,1,2,4$. The markers correspond
to single run simulations (N=6000) with no decay and learning rates $\eta=1$ (perceptron)
and $\eta=1.5$ (adatron). The solid lines are the results of 
numerical calculations
of the effective single pattern process with $M=20,000$  and time step 
$\Delta=0.05$.}
\end{figure}

Upon combining the results of the previous two paragraphs, we find a closed
set of exact equations relating the evolution of $R$, $D$ and
$G$ to the evolution of the local field distribution implied by
the measure in  (\ref{eq:Phi2}).
Setting $\theta_y=0$ in this equation, we find that the distribution
is generated by the following
stochastic process for a student-pattern overlap: 
\begin{equation}\label{eq:x}
  x_t=R_t y+ \sum_{t'}G_{tt'}f_{t'} + z_t +\theta_{xt},
\end{equation}
where $y$ and $z_t$ are independent Gaussian random variables with zero
mean and variances $\langle y^2\rangle=1$ and $\langle z_t z_{t'}
\rangle=D_{tt'}$. In general $x_t$ will depend on previous values of
$x$ via the term $[Gf]_t=\eta \sum_{t'} G_{tt'}k_{t'} F(x_{t'},y)$.

The evolution of the order parameters 
is given using the bare propagator $G_0$. To ensure
that the students' weight distribution
 will eventually reach a  
stationary state, we let the weights decay  with rate $
\gamma$. The bare propagator then takes the form
$[G_0^{-1}]_{tt'}=\delta_{t+1,t'}-\lambda \delta_{tt'}$, where
$\lambda\equiv 1-\Delta\gamma$. 
In the limit of $\Delta \rightarrow 0$, this corresponds to $[G_0]_{tt'}=\theta(t-t'-\Delta/2)\exp[-\Delta\gamma(t-t'-1)]$.
The equations (\ref{eq:GRD1}) and
(\ref{eq:hrhchG}) determine the evolution of the response function:
\begin{equation}\label{eq:evolutionG}
  [G_0^{-1}G]_{tt'}=\II_{tt'} + \alpha \sum_s 
  \langle \frac{\partial f_t}{\partial \theta_s} \rangle G_{st'},
\end{equation}
where now $\langle\cdot\rangle$ is the Gaussian averages over $y$ and
all $z_t$'s.
Using equation (\ref{eq:GRD1}) along with the relation
(\ref{eq:hr}), we find:
\begin{equation}\label{eq:evolutionR}
  [G_0^{-1}R]_t=\hr_t-\sum_s i\hG^T_{ts}R_s=\alpha \langle y f_t \rangle.
\end{equation}
The combination of equations (\ref{eq:GRD1})
 and (\ref{eq:hrhchG}) gives the evolution of $D$:
\begin{equation}\label{eq:evolutionD}
  [G_0^{-1}D]_{tt'}=\alpha\langle f_t [Gf+z]_{t'} \rangle
     =\alpha \langle f_t (x_{t'}-R_{t'}y)\rangle,
\end{equation}
where we have set $\theta_x$ to zero.
The evolution of the diagonal terms of $Q_t\equiv D_{tt}-R_t^2$
 can be implicitly calculated using
this equation, but the distribution of 
$x_{t+1}$ has to been known before $Q_{t+1}$
can be calculated. To avoid this difficulty, we use equation
(\ref{eq:evolutionD}) together with the scaling arguments  
$\hG_{tt'}=\mathcal{O}(\Delta^2)$ for $t<t'$, 
$\hG_{tt}=\mathcal{O}(\Delta)$ and $z_{t+1}-\lambda
z_t=\mathcal{O}(\Delta^{1/2})$ to determine for small $\Delta$ 
the evolution of $Q_t$ more explicitly:
\begin{eqnarray} \fl \label{eq:evolutionQ}
  Q_{t+1}=
  \lambda^2 Q_{t} 
  + \alpha\lambda\langle f_t x_t\rangle
  + \alpha \langle f_t x_{t+1}\rangle
  \nonumber
\\
 = \lambda^2 Q_{t} 
  +2\alpha\lambda\langle f_t x_t \rangle
  +\alpha\langle f_t([z+Gf]_{t+1}-\lambda[z+Gf]_t) \rangle \nonumber
  \\
 =\lambda^2 Q_{t} 
  +2\alpha \lambda \langle f_t x_t \rangle
  + \alpha \langle f_t^2 \rangle 
  + \mathcal{O}(\Delta^{3/2})\qquad (\Delta\rightarrow 0)
\end{eqnarray}
The equations (\ref{eq:evolutionG})-(\ref{eq:evolutionQ}) could perhaps have
been found using less sophisticated methods, but the strength of the
generating functional method is that it is also capable of producing
 the joint local field
distribution $P(x,y)$ generated by (\ref{eq:x}).
The generalization error is a direct function of these order parameters,
while the training error is a slave of the local field distribution
governed by them:
\begin{equation}
  E_g(t)=\frac{1}{\pi}\arccos\left(\frac{R_t}{\sqrt{Q_t}}\right),\qquad
  E_t(t)=\langle \theta(-x_t y)\rangle
\end{equation}
The evolution of the order parameters can be calculated numerically by
a Monte Carlo procedure similar to the single spin
procedure outlined in \cite{EissOppe92}. 
The general idea is to follow
the evolution of $M$ patterns overlaps. For each of these patterns,
one generates at time $t=0$ a teacher overlap $y$ from the standard
normal distribution.  
Time is discretized with unit $\Delta$. At each time step and for
\emph{each} pattern, one
generates the Gaussian noise $z_t$, correlated with
the previous noise values $z_{t'}$ \emph{for that particular pattern} and
a Poissonian random variable $k_t$. Averages over all patterns are
Monte Carlo implementations of the averages occuring in the
evolution equations for $D$,$R$ and $G$. By increasing $M$ and
decreasing $\Delta$ the evolution of the
$N\rightarrow\infty$-perceptron can be calculated up to arbitrary precision.
This is shown for various $\alpha$ in figure
\ref{fig:evolution} with $M=20,000$ and $\Delta=0.05$.
The figures illustrate that the  agreement of the theory with the
simulations is quite satisfactory.

\subsection{Batch learning}
So far, we have treated only the case of on-line learning. This is the
most widely applied learning scenario, but much of the analytical work
on learning with restricted training sets has been devoted to
off-line or batch learning. In batch learning one first calculates the
average effect of learning (a large sample of) the entire training
set, before making a weight update. For  small learning rates, batch
and on-line learning ought to generate
 the same macroscopic flow. For completeness
we discuss here what changes when we switch from an on-line to a batch
scenario. The effect on the theory as presented above is the
disappearance of the extra noise term $\langle f_t^2 \rangle$ in the
evolution of $Q$ in equation (\ref{eq:evolutionQ})
and the replacement of the Poisson variable $k_t$ by its average
$\Delta/\alpha$. 
The intuition behind the first change is that big
changes in the student weight vector can no longer 
happen after a single pattern is
presented; the weights undergo a much smoother
evolution due to the averaging of the update over all patterns. 
As a result of the second change, the student training pattern overlap
becomes:
\begin{equation}
  x_t=R_t y +z_t + \eta \frac{\Delta}{\alpha}
   \sum_s G_{ts} F(x_s,y)
\end{equation}
This equation was derived earlier in the context of gradient-descent
batch learning by Wong et. al
using an elegant application of the
 dynamical cavity method  \cite{WongLiLuo00}.
Again, the reason for the
change in a training pattern overlap $x_t$ is that instead of big
changes when $k_t$ times that particular pattern is presented to the
student in time interval $(\Delta t,\Delta (t+1))$, now during an interval
interval $x_t$ feels the average effect of the influence of the pattern.
These are the only changes necessary in the present analysis when
 switching from on-line
to batch learning.

\subsection{Linear learning rules}
The average occurring in the evolution of the response function G
in (\ref{eq:evolutionG}) can be explicitly calculated 
if the student is using a learning rule that is linear in $x$,
e.g. the linear, Hebbian or adaline rules. For this type of rules
of
 the form $F(x,y)=g(y)-c x$, we find:
\begin{equation}
  \frac{\partial f_t}{\partial x_{t'}} 
  =
  - \eta c k_t \left(
    \delta_{tt'}+\sum_s G_{ts}\frac{\partial f_s}{\partial x_{t'}}
  \right).
\end{equation}
To causality of $G$ allows us to perform the Poisson averages and
a little matrix algebra leads to
\begin{equation}
  G_0^{-1}G=
  \II-\Delta c \eta \left[\II+c\eta\frac{\Delta}{\alpha}G\right]^{-1}G
\end{equation}
The resulting response is translation invariant,
i.e. $G_{tt'}=G(\Delta(t-t'))$ for $t\geq t'$. The on-line response
found here for linear learning rules agrees with the batch-results
found for the linear rule in
\cite{HertKrogThor89} and the adaline rule in \cite{WongLiTong00}. The
Fourier transform of the previous relation reads
\begin{equation}
  G^{-1}(\omega)=\gamma - i\omega 
    + c\eta\frac{1}{1+\frac{c \eta}{\alpha} G(\omega)}
\end{equation}
This equation is analysed in \cite{HertKrogThor89}. For
$\gamma=0$ and for $c\not=0$, a transition in the behaviour of the
response takes place at $\alpha_c=1$. This position is
identical for on-line and off-line learning. 
The nature of this transition is easily understood. Without decay, the
evolution of the weight vector is confined to the linear subspace
spanned by the patterns in the training set. 
Below $\alpha_c=1$, the random patterns are
unlikely to span the whole $N$-dimensional space, resulting in a
non-decaying part of the response function. This argument is valid for
general rules without decay.

The student overlap with a particular pattern 
can also be written in a more explicit way:
\begin{equation}
  x=\left[\II+c\eta GK\right]^{-1}(Ry+z+\eta g(y) Gk),\quad
  K_{tt'}\equiv k_t \delta_{tt'}
\end{equation}
The final results are rather cumbersome, but all the averages appearing
in the evolution of the order parameters involving
$k$ and $z$ can be done without any problems. The only remaining
integrals are of the form 
$\langle g(y) y\rangle$ and $\langle g(y)^2 \rangle$ with the
standard Gaussian measure.

\subsection{Infinite training sets}
To compare our results to the well-known unrestricted training set
results, we take the limit $\alpha\rightarrow\infty$. In this case
the probability of repeating an example is zero. This is reflected in the
fact that $\langle k\rangle \rightarrow 0$ as
$\alpha\rightarrow\infty$. 
Given $y$, the local
fields $x$ are random variables given by:
\begin{equation}
  x_t=yR_t + z_t,
\end{equation}
or, equivalently, $x_t$ is a Gaussian random variable with mean $yR_t$
and covariance $D_{tt'}$. The effects of the
retarded self-interaction caused by $G$ thus completely vanish. If we
go to a continuum time description, we recover equations found in 
e.g. \cite{KinoCati92,BiehSchw92}. The
evolution of the student-teacher overlap and the student self-overlap
 are given by
\begin{eqnarray}
  \frac{dR}{dt}=-\gamma R_t + \eta \langle y F(x,y) \rangle_{t}
   \\ 
  \frac{dQ}{dt}=-2\gamma Q + 2 \eta \langle x F(x,y) \rangle_{t} +
    \eta^2 \langle F(x,y)^2 \rangle_{t},
\end{eqnarray} 
with the Gaussian single-time 
average defined by $\langle x \rangle_t=0, \langle y
\rangle_t =0, \langle x^2\rangle_t=Q_t, \langle y^2 \rangle_t=1,\langle xy
\rangle_t =R_t$.

\section{Stationary state}\label{sec:statstate}

\begin{figure}
  \epsfig{file=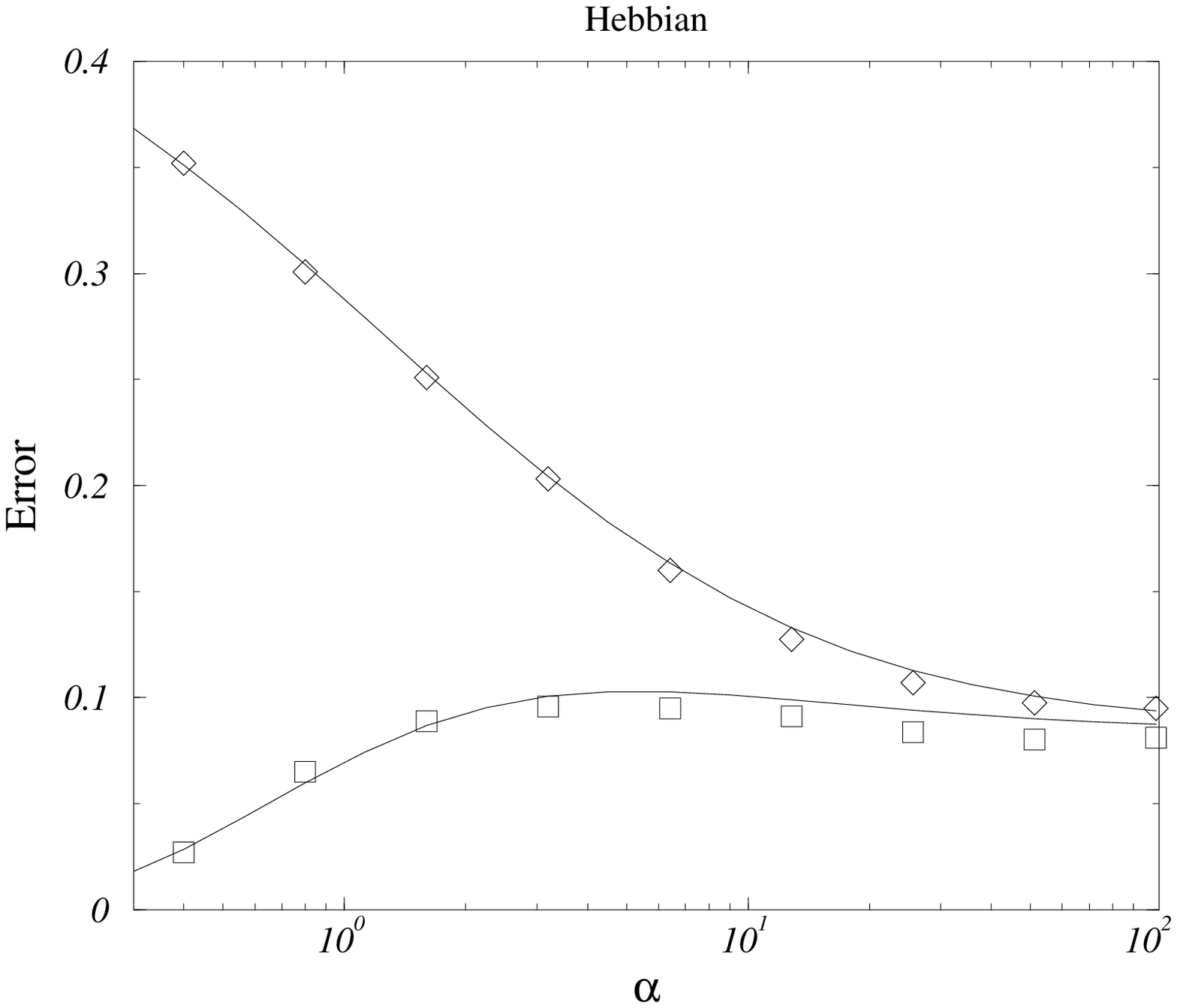,width=0.48\textwidth}
  \epsfig{file=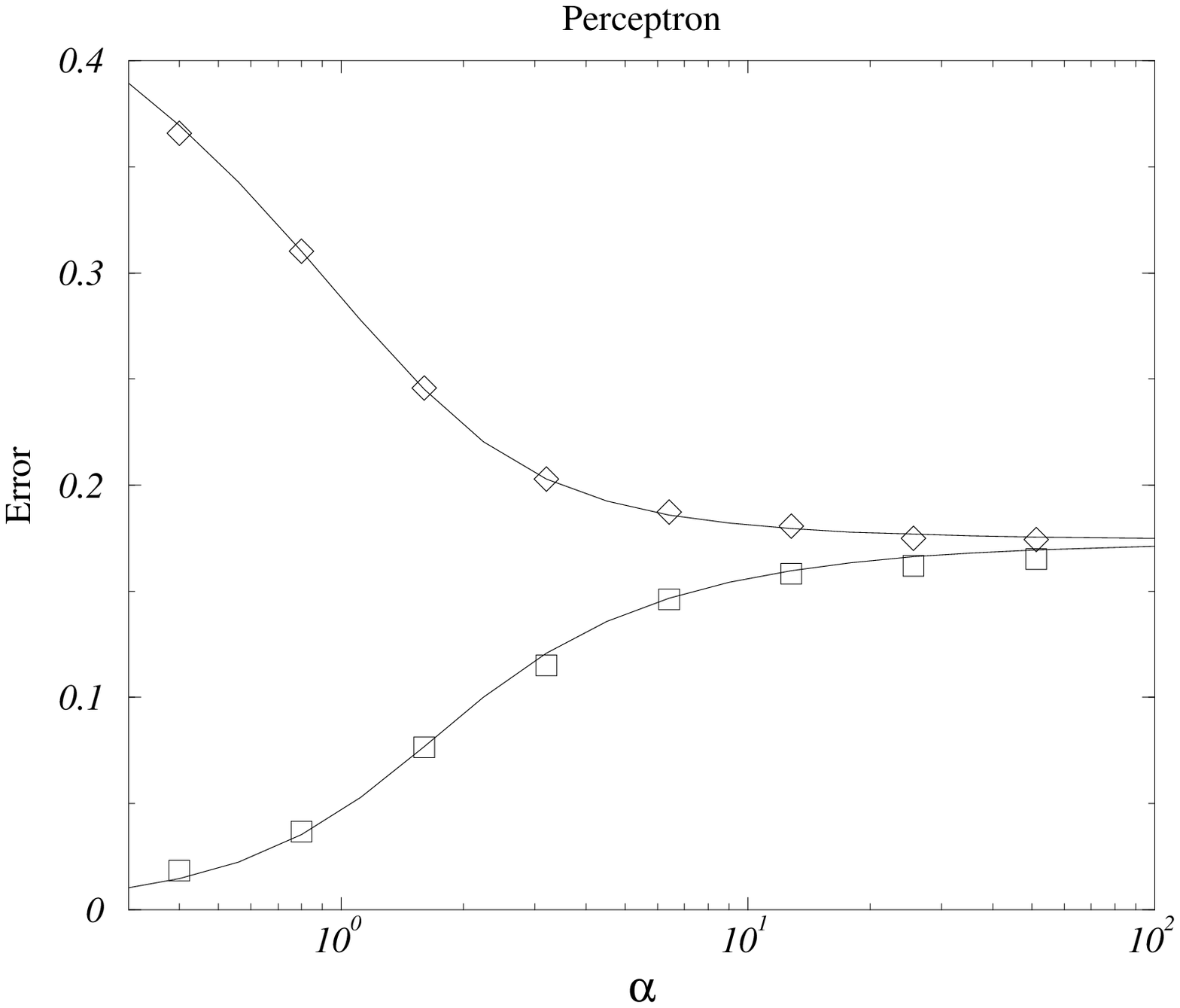,width=0.48\textwidth}

\caption{\label{fig:statalpha}Stationary generalization (upper line) and training
error (lower line) for perceptron (left) and a Hebbian (right) 
learning rules. Learning
rate $\eta=1$ and decay $\gamma=0.1$. Markers are simulation results
of a single run with $N=6000$ inputchannels. Solid lines are
theoretical predictions, obtained by solving  (\ref{eq:x}) }
\end{figure}

Many learning rules will not reach a stationary state that is
independent of the initial conditions, as soon as weight decay is
absent.
Weight decay, or another type of constraint, may also  be
necessary to bound the length of the student vector.
In the
Hebbian case, for example, the student weights keep on growing in the
direction of the perceived teacher, regardless
of the size of the training error. 
In order for the student to reach a stationary state, we 
assume that
the weight decay $\gamma$ is large enough to bound $R$
and $C$ and that the integrated response or susceptibility $g$ is finite:
\begin{equation}
  g\equiv \lim_{t\rightarrow\infty}\Delta \sum_{t'} G_{tt'} < \infty.
\end{equation}
This condition, known in the disordered systems literature as
\emph{absence of anomalous response}. We also assume that for 
sufficiently large $t$ the order parameters become time translation
invariant: $R_t=R$, $G_{t+\tau,t}=G_\tau$, $D_{t+\tau,t}=D_\tau$. 
These assumptions are related to the replica symmetry ansatz in the
replica equilibrium analyses \cite{MezaPariVira87}. 
We
split the covariance kernel $D_t$ in a persistent part 
$d=\lim_{t\rightarrow\infty}D_t$ and a non-persistent part
$\tilde{D}_t=D_t-d$. 
If $d$ exists, then 
\[
  d=\overline{D}\equiv\lim_{T\rightarrow\infty}\frac{1}{T} \sum_{t=0}^T D_t
\]
Given time
translation invariance, one derives from equations (\ref{eq:evolutionR})
and (\ref{eq:evolutionD}) that
\begin{eqnarray} \label{eq:statRd}
  R=\frac{\eta}{\gamma}\langle y F(x,y) \rangle 
  \\
  d=\lim_{\tau\rightarrow\infty}
    \lim_{t\rightarrow\infty}
    \frac{\eta}{\gamma}\langle F(x_{t+\tau},y) (x_t-yR) \rangle
  =    
  \frac{\eta}{\gamma}\langle \overline{F} (\overline{x}-yR) \rangle
\end{eqnarray}
A relation earlier found  involving the covariance (\ref{eq:GRD1}) now
 yields
\begin{equation}\label{eq:statd2}
  d=\alpha \overline{ G \langle f f^T \rangle G^T} 
  = \frac{\eta^2g^2}{\alpha} \langle \overline{F}^2  \rangle,
\end{equation}
while the stationary value of $Q$ can be found from (\ref{eq:evolutionQ}):
\begin{equation}\label{eq:statQ}
  Q=\frac{\eta}{\gamma}\langle F(x,y)x \rangle +
   \frac{\eta^2}{2\gamma}\langle F(x,y)^2 \rangle.
\end{equation}

\begin{figure}
  \centerline{\epsfig{file=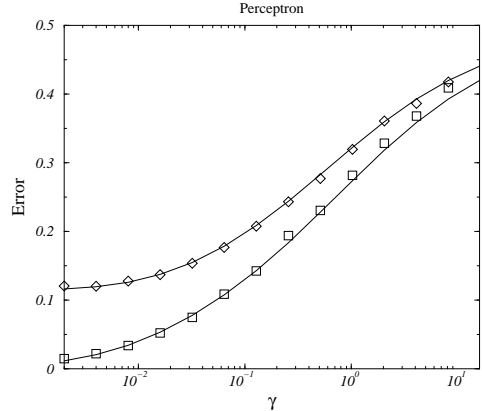,width=0.48\textwidth}}
\caption{\label{fig:statdecay}Stationary generalization (upper line) and training
error (lower line) for perceptron learning rule with various decay
rates $\gamma$. Training set size $\alpha=4$.}
\end{figure}

All the averages either involve a single time or two infinitely separated
distant times. We lack an explicit expression for the 
single time probability distribution
of $x_t$. The probability of $x_t$ is
related to realisations of $x$ at previous times via the response
 function $G$. This makes the
evaluation of the averages as hard as solving the dynamical
equations themselves. The  same problem exists in the field of (Ising) spin
glasses and recurrent neural networks. In those cases where the
stationary state is in detailed balance (e.g. when the dynamics are of
gradient descent type and the systems feels a Gaussian white noise) 
a fluctuation dissipation relation connects the
correlation $C$ and the response $G$. It is known for such systems 
that when 
calculating the persistent and single time parts of the correlation and
the integrated response, the non-persistent parts can be chosen
arbitrarily as long as the FDT is obeyed. In particular one can set
them to zero 
and take only the persistent parts and 
the integrated response into account. 
Although there are big differences between the learning
perceptron discussed here and the aforementioned spin systems 
(for one, the learning rule $F$ does not have to be a gradient),
 we assume this decoupling property of persistent from non-persistent
parts still holds\footnote{Note that a rigorous proof would first
require the derivation of a non-equilibrium generalization of FDT
theorems.}.
 We replace the equilibrium 
distribution of $x$ generated
by equation (\ref{eq:x}) by a distribution generated by a stochastic
relation containing only the integrated response $g$ and random
variables described by the persistent
part of the covariance matrix $D$, the single-time correlation $Q$ and
the student-teacher overlap  $R$:

\begin{equation} \label{eq:statxt}
  x_t =y R + z + \tilde{z}_t + \frac{\eta}{\alpha}g\overline{F},
\end{equation}
where $y,z$ and $\tilde{z}_t$ are all independent Gaussian random
variables with zero mean and covariances $\langle y^2 \rangle
=1$,$\langle z ^2 \rangle =d $ and 
$\langle \tilde{z}_t
\tilde{z}_{t'}\rangle=(Q-R^2-d)\delta_{tt'}$. The average learning
term $\overline{F}$ for a specific pattern with a certain
$(y,z)$, can be expressed self-consistently as:
\begin{equation}\label{eq:overlineF}
  \overline{F}_{yz}\equiv
  \lim_{T\rightarrow\infty}
  \frac{1}{T}\sum_{t<T}F(x_t,y)
  =
  \int\! d\tilde{z}\,p(\tilde{z}) F(yR+z+\tilde{z}+
   \frac{\eta g}{\alpha}\overline{F}_{yz},y )
\end{equation}
For Hebbian learning one has $\overline{F}=\sgn(y)$, but in general
(\ref{eq:overlineF}) will be a
transcedental equation, so one has to revert to numerical
methods to solve it. Once $\overline{F}$ can be found for
any point $(y,z)$, the remaining two independent Gaussian integrals
over $y$ and $z$ can be evaluated to close equations
(\ref{eq:statRd}) to (\ref{eq:statQ}).
The remaining closed set can be solved numerically. Results for
Hebbian and perceptron learning rules and various training sets sizes 
are presented in
figure \ref{fig:statalpha}. For the perceptron rule, the results shown
in figure \ref{fig:statdecay} compare $E_g$ and $E_t$ for different
decay strengths. Perceptron results are independent of $\eta$.
The theoretical predictions seem to be in
almost perfect agreement with the simulations. Although 
no adatron results are
shown, we expect that the proposed procedure is equally valid for the
this latter rule. Our method of calculation is only valid when $G$
is time translation invariant and the integrated response is bounded. For this
to happen, we need the presence of a weight decay. The complication is
that any
decay, however small, 
 will cause the adatron student weight vector to vanish. An
alternative way of ensuring that the student ensemble reaches a
stationary state that does not exhibit this problem is
by constraining $\bsigma$ to a sphere. This can be implemented by
choosing $\gamma_t \propto (Q_t-1)$.  However, the
adatron rule yields zero training error in this setup. This causes
other problems in numerically evaluating this stationary state
equations.

\subsection{Distribution of local fields}

\begin{figure}
\begin{center}
\begin{tabular}{rcc}
\parbox[b][40mm][c]{6pt}{$y$}
& 
\epsfig{file=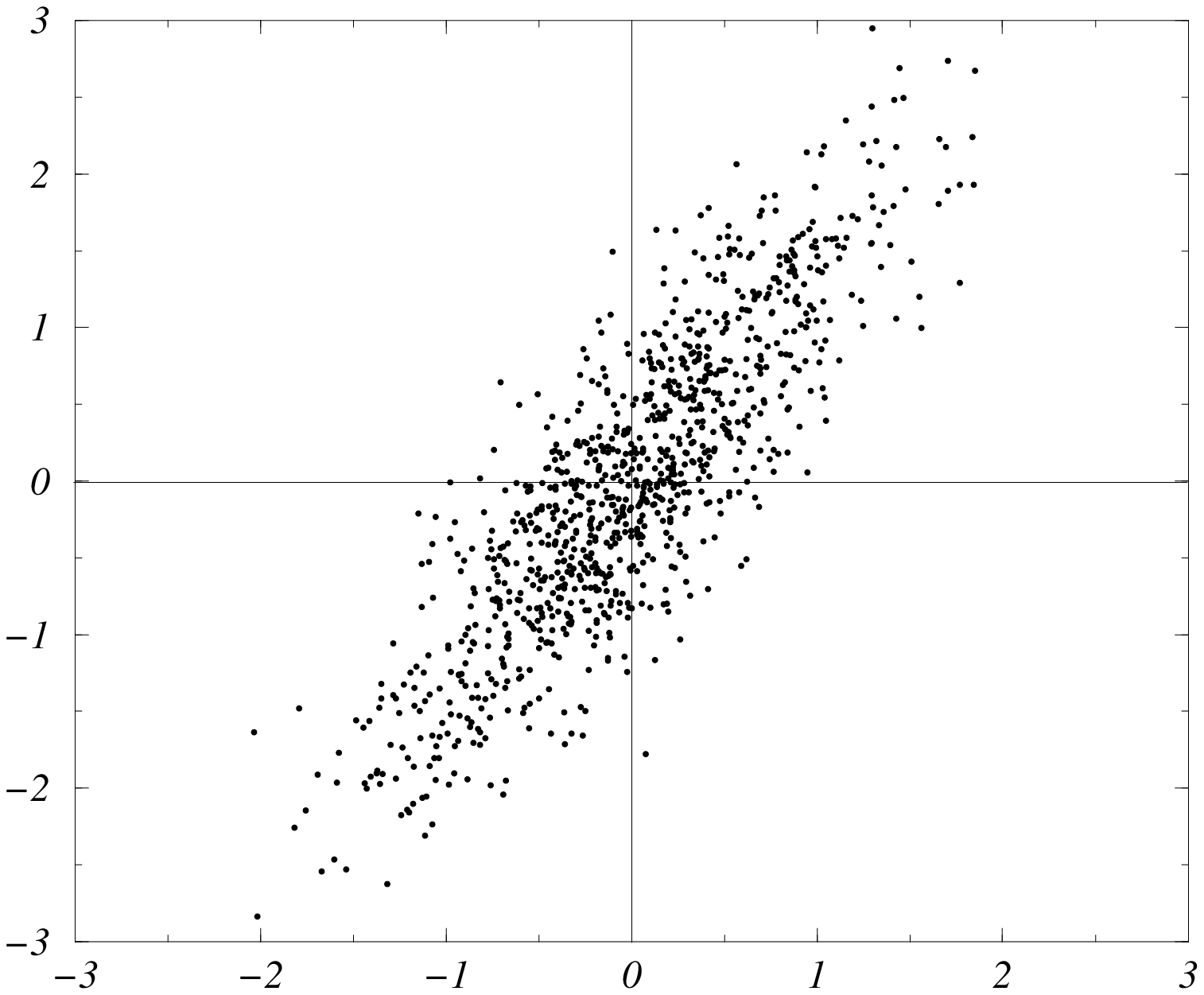,width=40mm,height=40mm}
&
\epsfig{file=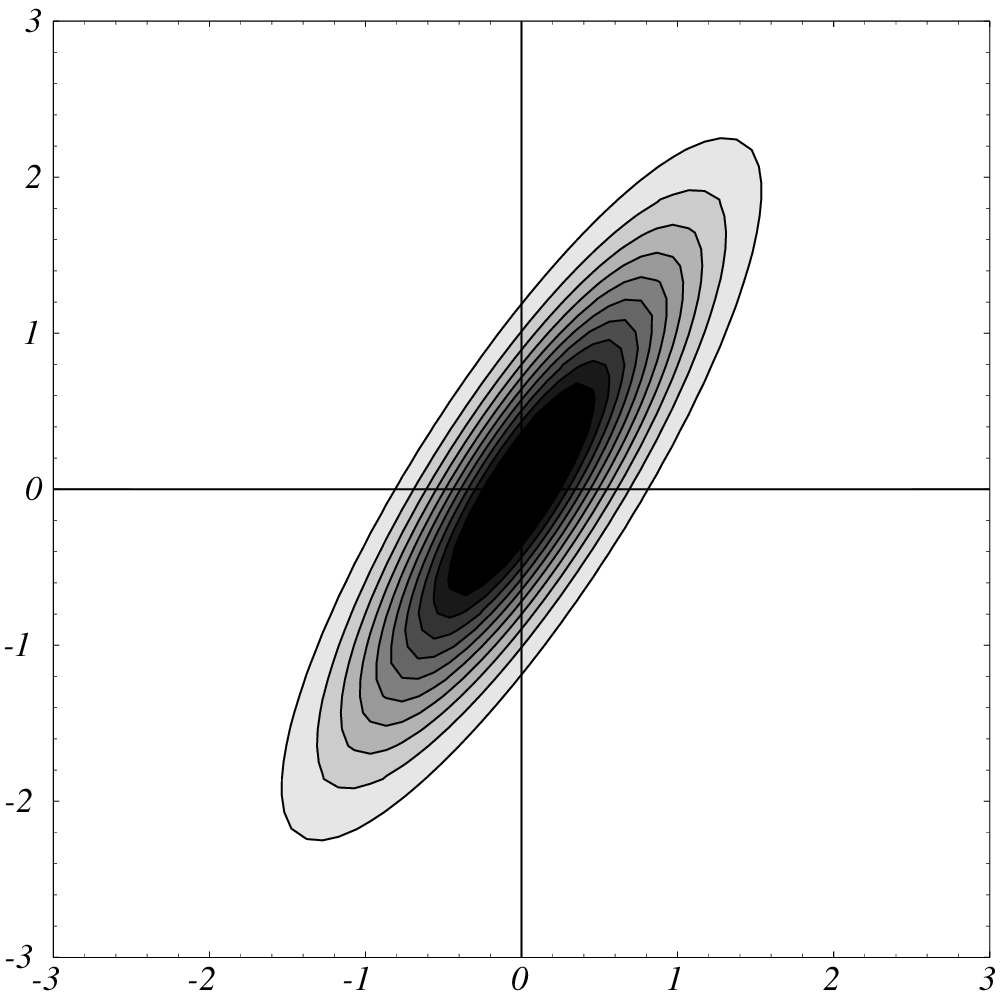,height=40mm}
\\
\parbox[b][40mm][c]{6pt}{$y$}
&
\epsfig{file=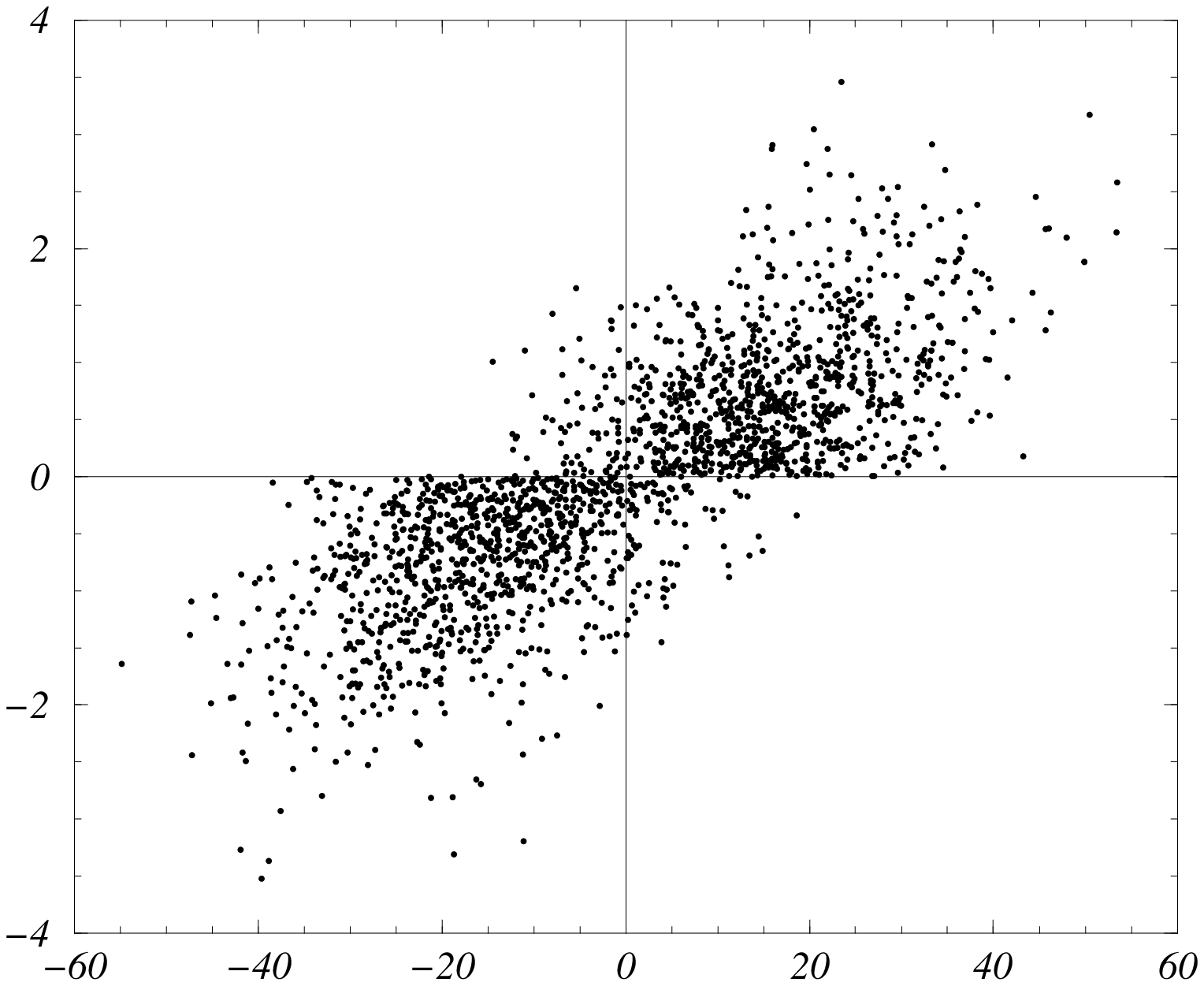,width=40mm,height=40mm}
&
\epsfig{file=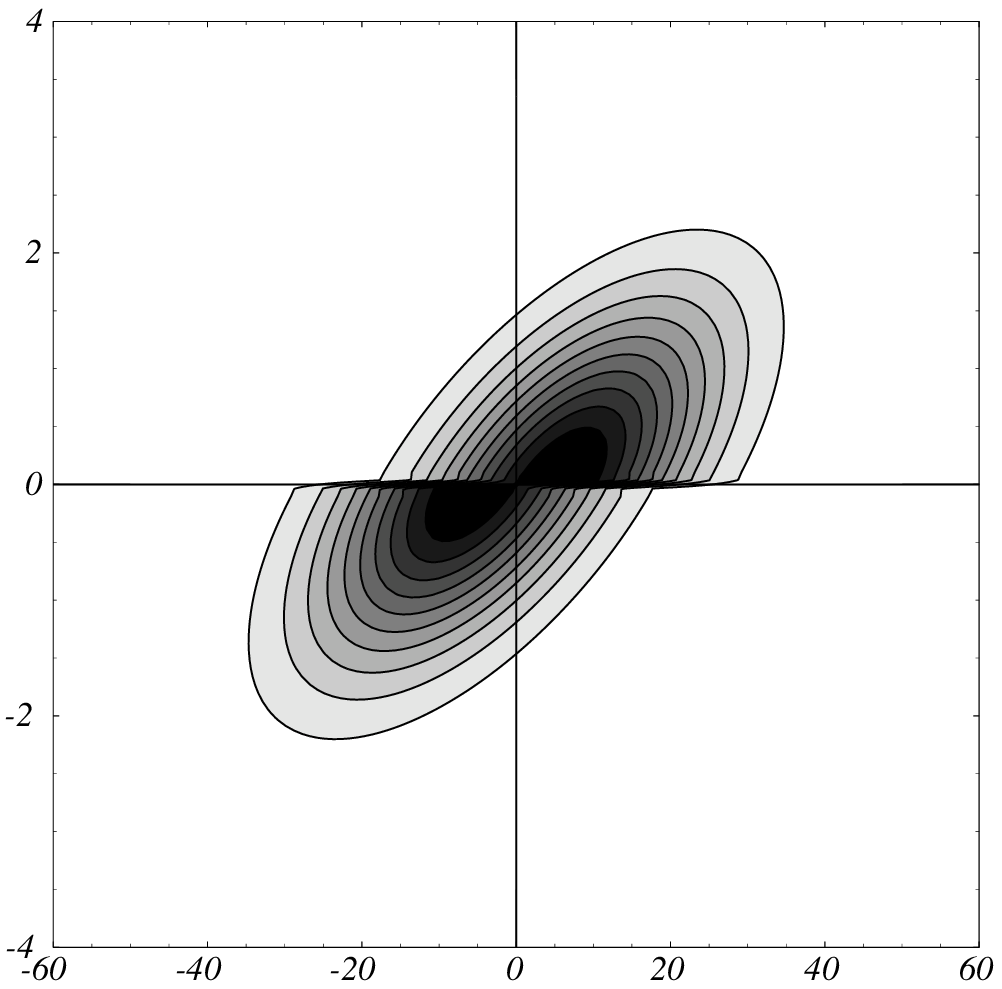,height=40mm}
\\
\parbox[b][40mm][c]{6pt}{$y$}
&
\epsfig{file=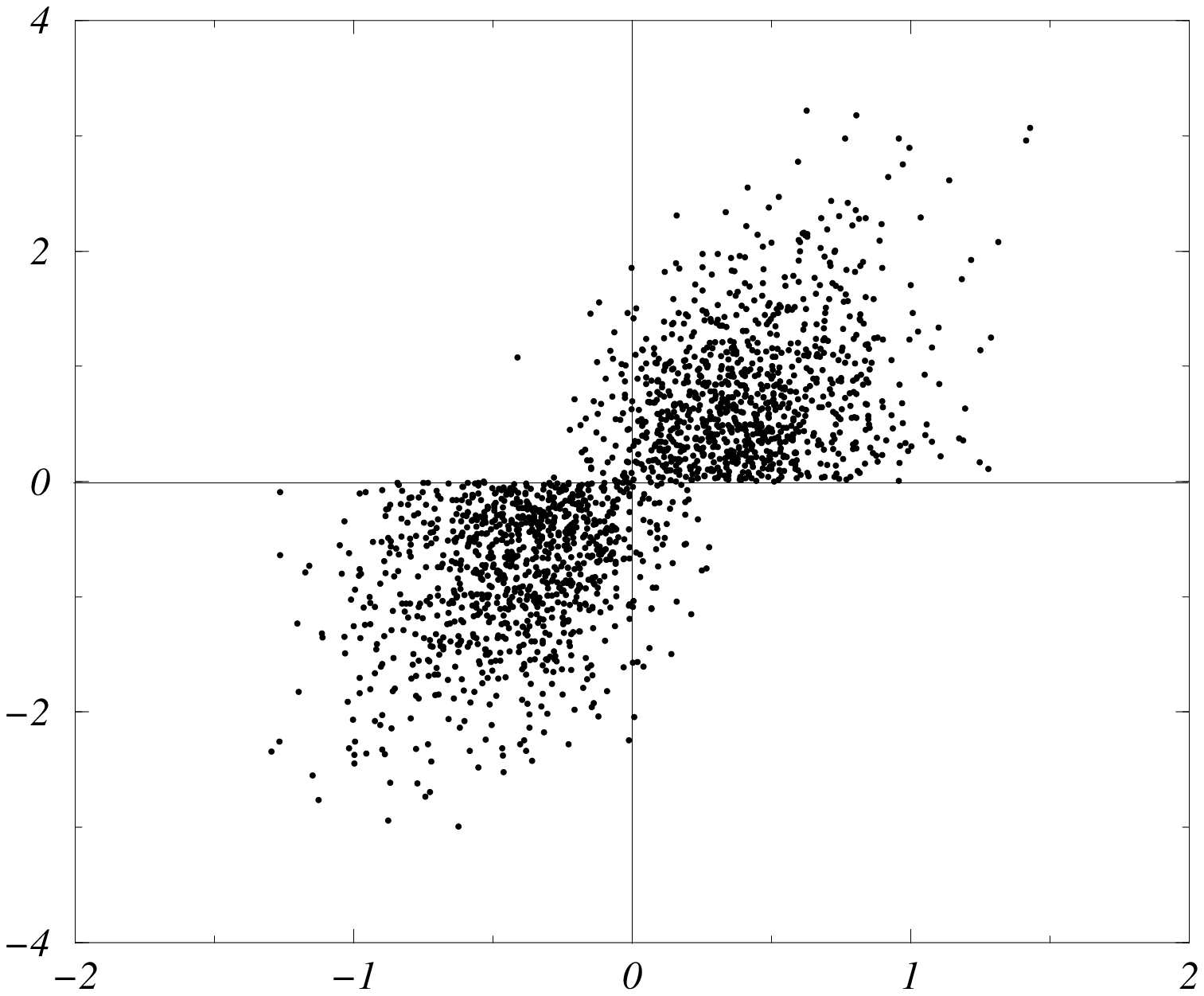,width=40mm,height=40mm}
&
\epsfig{file=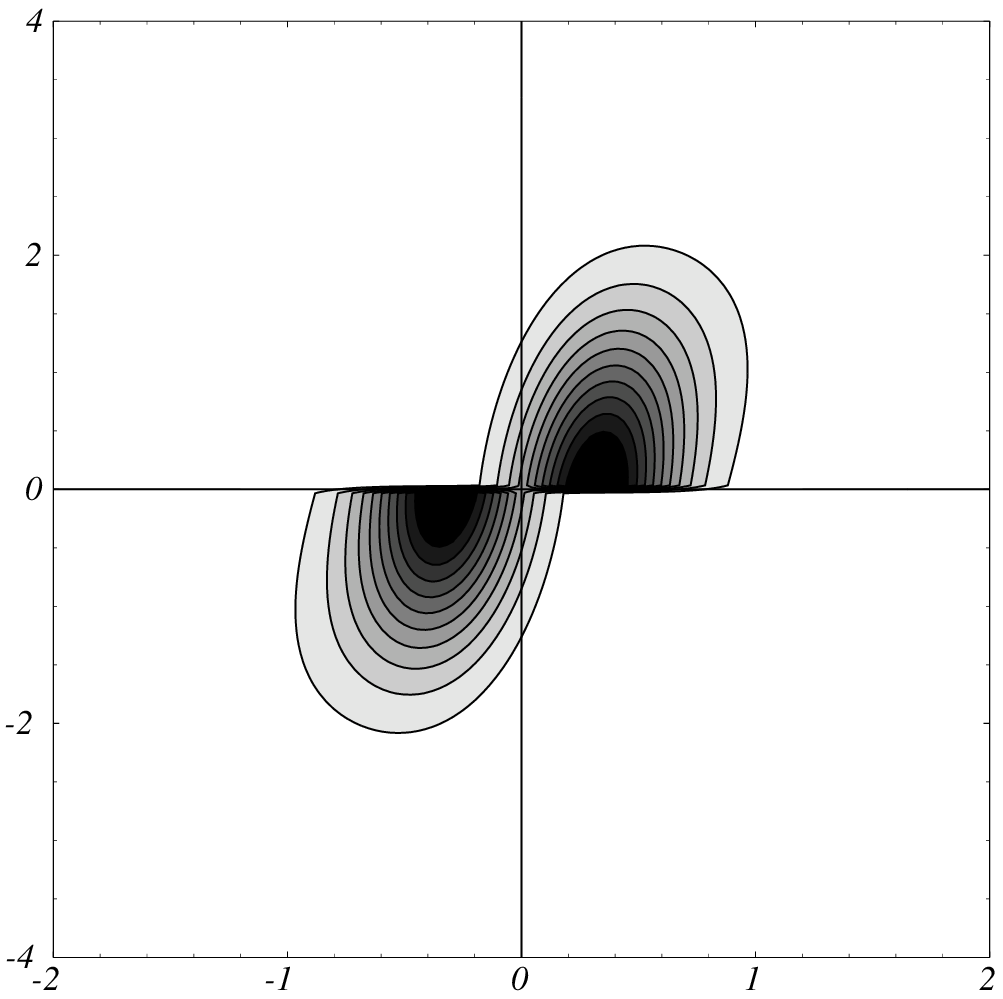,height=40mm}\\
 & ~~x &  x
\end{tabular}
\end{center}

\caption{\label{fig:localfield} Stationary local fields distributions
$p(x,y)$ for an infinite ($\alpha=\infty$)
training set after perceptron learning (top), and two finite
($\alpha=1$) training sets after Hebbian learning (middle), or
perceptron (bottom) learning.  Left are simulation results, right are
theoretical predictions in the form of contour plots.  The infinite
training set yields a joint Gaussian distribution, Hebbian learns
gives only a conditionally Gaussian $p(x|y)$ and the perceptron rule
deviates even further from the Gaussian shape. Learning rates are
$\eta=1$ and decay coefficients are $\gamma=0.1$ in all three graphs.}
\end{figure}

As seen earlier, a big  simplifying effect of the limit
$\alpha\rightarrow\infty$ is to render the local fields $x$ and $y$
Gaussian. This happens irrespective of the learning rule involved. As
soon as $\alpha<\infty$, the effect of the extra term $Gf$ in equation
(\ref{eq:statxt}) sets in and the Gaussian form of the distribution
evaporates for non-linear rules.  The non-Gaussian form of the joint
local field distribution has been discussed at length in
\cite{CoolSaad00}, but equation (\ref{eq:statxt}) gives an intuitive
idea of the origin of the deviations reported there.

For a Hebbian learning rule, $F(x,y)=\sgn(y)$, the conditional
distribution $p(x|y)$ remains Gaussian with variance D, but will be
shifted away from the mean $yR$ by the amount $\eta g
\sgn(y)/\alpha$. An example with $\alpha=1$ and $\gamma=0.1$ is shown
in figure \ref{fig:localfield}b. For the perceptron learning rule,
this is no longer true. The random variables $y$ and $z$ are
independently distributed Gaussian variables.  From
(\ref{eq:overlineF}) we find that:
\begin{equation}\label{eq:overlineFperc}
  \overline{F}= 
  \frac{1}{2}\sgn(y)
  -\frac{1}{2}\erf\left(
    \frac{ yR+ z+\frac{\eta g}{\alpha}\overline{F}}
         {\sqrt{2(D-d)}}
  \right)
\end{equation}

\begin{figure}
\begin{center}
\begin{tabular}{r@{}cr@{}c}
&$\overline{F}$& & \hspace{3cm}$\overline{x}$ \\
\parbox[b][40mm][c]{8pt}{$z$}
&  
\epsfig{file=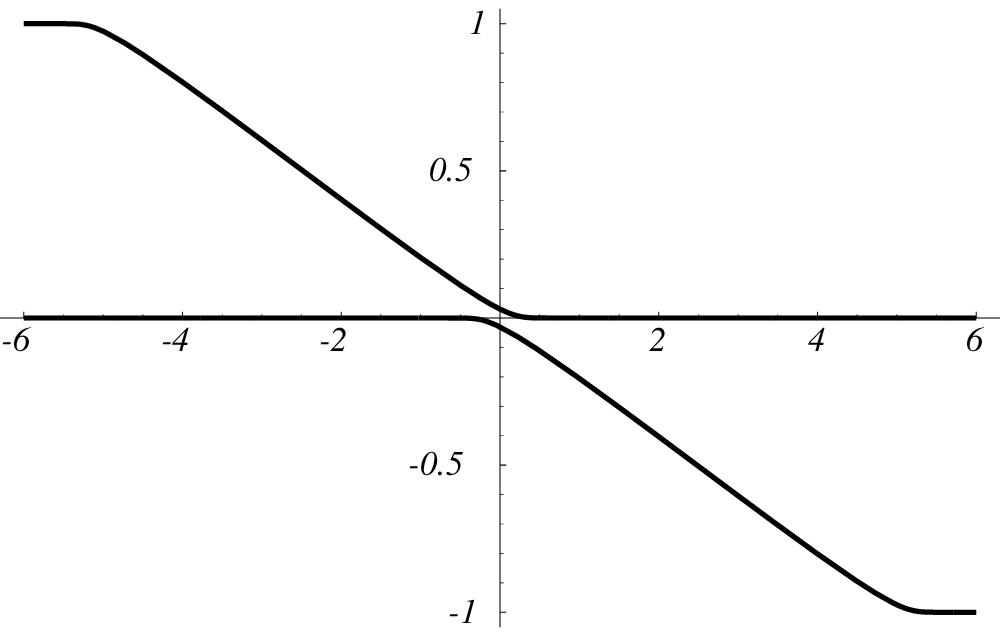,width=40mm,height=40mm}
&
\parbox[b][31mm][c]{8pt}{$z$}
&
\epsfig{file=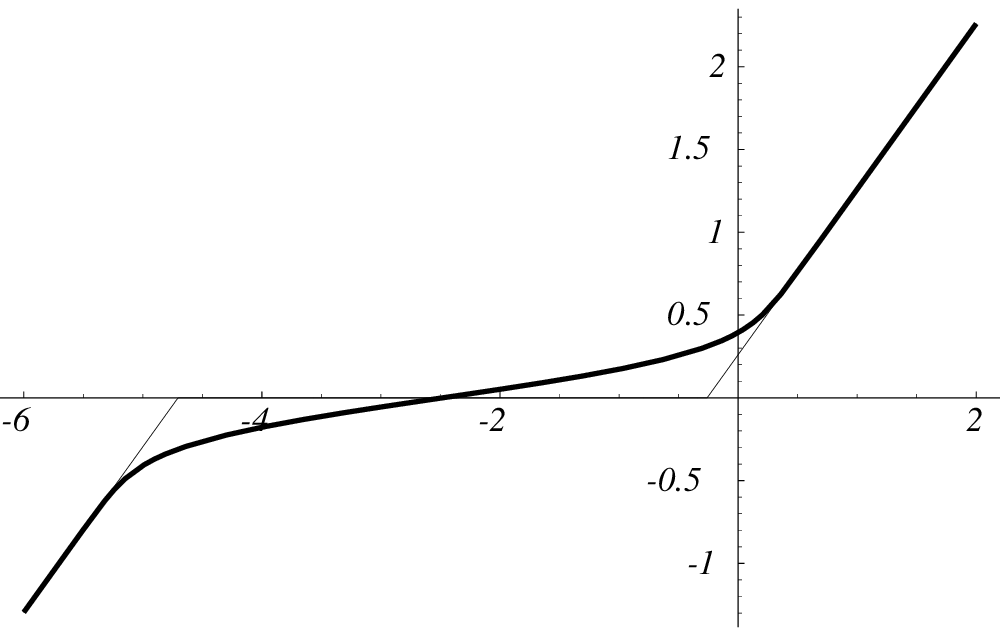,height=40mm}
\end{tabular}
\end{center}

\caption{\label{fig:F}a) The relationship between $\overline{F}$ (see
equation (\ref{eq:overlineFperc})) and
$z$ for $y=1$ (upper line) and $y=-1$ (lower line) for the stationary
state of a perceptron with $\alpha=1$ and $\gamma=0.1$. The width of
the sloping part is close to $\eta g/\alpha$, the abcissas are near
$-yR$. b) The time average of $x_t$ as function of $z$, given
$y=1$. The abcessas of the thin straight lines 
 are near $-yR-\eta g/\alpha$ and $-yR$. Due to the
Gaussian measure of $z$ centered at the origin, the part close
to the x-axis is the main important contribution.}
\end{figure}

Samples of the $(y,z)$ statistics for $\alpha=1$ and $\gamma=0.1$ of
$\overline{F}$ as a function of $z$ are shown in figure \ref{fig:F}a
for $y>0$ (top) and $y<0$ (bottom). The width of the sloping segment
is $\eta g/\alpha$, while the size of $\sqrt{D-d}$ determines the
rounding at the edges.  The value of $\overline{x}$ corresponding to $y=1$ as a
function of the Gaussian disorder $z$ is drawn in figure
\ref{fig:F}b. For $y$ positive and roughly $z>-yR$, one has
$\overline{x}=yR+z$, whereas for $z<-yR-\eta g/\alpha$ one finds
$\overline{x}=yR+z+\eta g/\alpha$. For $z$ in the range $-yR-\eta g/\alpha<z<-yR$, we
find $\overline{x}\approx 0$. In this particular example (using the
same values for the order parameters as the graphs shown in figure
\ref{fig:localfield}c) $\sqrt{d}\approx 0.27$ so that the Gaussian
measure confines $z$ close to the origin.  Thus the resulting local
field distribution is distinctly non-Gaussian as shown in figure
\ref{fig:localfield}c.

\section{Conclusion}

In this paper, we have studied the statics and dynamics of an ensemble
of students learning on-line the classification of a large number of
examples. This problem boils down to solving a large number of coupled
stochastic difference equations, each corresponding to a single input
channel. The situation is complicated by the existence of disorder in
the form of the composition of the training set. Using the generating
function method we have transformed this Markovian system of $N$
coupled equations in the limit of $N$ to infinity into an effective
single pattern process. The price paid for this reduction is that the
new process has noise which is correlated in time and the presence of 
a retarded
self-interaction in the system, which make the dynamics
non-Markovian. In principle it is possible to calculate the evolution
of the system analytically, but in general it will be impossible to
pursue this after the very first few time steps. However, the process
can be solved numerically up to arbitrary precision.

Our calculation provides a solid basis for the further
 analytical study of linear
rules. For non-linear rules the importance of our exact macroscopic
 dynamical
equations is  mainly in the insight they can give into the
behaviour of different learning rules and the possibility they create to
study and solve stationary states of both on-line and batch, gradient and
non-gradient learning. Until now, the stationary states of these kinds
of learning processes have only been directly accessable with tools from
equilibrium statistical mechanics, requiring detailed balance. This
confined  analyses to batch gradient-descent learning. This
restriction has now been lifted. From our macroscopic
 evolution equations we can
extract the stationary state equations very easily if we assume
 time translation invariance and the absence of anomalous response. 
We have not yet addressed the issue
where this is likely to hold for on-line learning. To reduce the
time-dependent order parameters like the student-autocorrelation and
the student-response to a finite set of scalar order parameters, we
apply a method we know from similar spin-glass problems based on the
detachment of single-time and persistent order parameters from the
non-persistent ones. The procedure
consists of removing all non-persistent parts of the order parameters
 (except for the single time quantities), retaining only a small
closed  set of equations containing just four (Q,R,d,g) scalar
macroscopic order parameters. Whether this last 
procedure is indeed exact, remains to be seen and will be the subject
of a future study, but the numerical
evidence clearly suggests that the underlying assumption holds.


\section*{References}

\end{document}